\documentclass[twocolumn]{aastex631}
\usepackage{CJK}
\usepackage{color, amssymb, amsmath, natbib, color}
\usepackage{multirow, placeins, subfigure}
\usepackage{graphicx}
\usepackage{epstopdf}
\usepackage{bm}
\usepackage{overpic}
\usepackage{hyperref}
\usepackage{natbib}
\usepackage{enumitem}

\newcommand{\Msunh}{\>h^{-1}\rm M_\odot}

\newcommand{\Gpch}{\>h^{-1}{\rm {Gpc}}}
\newcommand{\Mpch}{\>h^{-1}{\rm {Mpc}}}

\definecolor{myblue}{RGB}{0,0,255}

\begin{document}
\begin{CJK*}{UTF8}{gbsn}

\title{Exploring the signature of assembly bias and modified gravity using small-scale clusterings of galaxies \\ }

\correspondingauthor{Zhongxu Zhai, Xiaohu Yang}
\email{zhongxuzhai@sjtu.edu.cn, xyang@sjtu.edu.cn}

\author[0000-0003-3203-3299]{Yirong Wang (王艺蓉)}
\affiliation{State Key Laboratory of Dark Matter Physics, Tsung-Dao Lee Institute \& School of Physics and Astronomy, Shanghai Jiao Tong University, Shanghai 200240, China}
\affiliation{Shanghai Key Laboratory for Particle Physics and Cosmology, and Key Laboratory for Particle Physics, Astrophysics and Cosmology, Ministry of Education, Shanghai Jiao Tong University, Shanghai 200240, China}

\author[0000-0001-7984-5476]{Zhongxu Zhai (翟忠旭)}
\affiliation{State Key Laboratory of Dark Matter Physics, Tsung-Dao Lee Institute \& School of Physics and Astronomy, Shanghai Jiao Tong University, Shanghai 200240, China}
\affiliation{Shanghai Key Laboratory for Particle Physics and Cosmology, and Key Laboratory for Particle Physics, Astrophysics and Cosmology, Ministry of Education, Shanghai Jiao Tong University, Shanghai 200240, China}

\author[0000-0003-3997-4606]{Xiaohu Yang (杨小虎)}
\affiliation{State Key Laboratory of Dark Matter Physics, Tsung-Dao Lee Institute \& School of Physics and Astronomy, Shanghai Jiao Tong University, Shanghai 200240, China}
\affiliation{Shanghai Key Laboratory for Particle Physics and Cosmology, and Key Laboratory for Particle Physics, Astrophysics and Cosmology, Ministry of Education, Shanghai Jiao Tong University, Shanghai 200240, China}

\author[0000-0003-3578-6149]{Jeremy L. Tinker}
\affiliation{Center for Cosmology and Particle Physics, Department of Physics, New York University, 726 Broadway, New York, NY 10003, USA}

\begin{abstract}
We apply a halo velocity bias model, $\gamma_{f}$, within the Aemulus simulation suite for General Relativity (GR) to investigate its efficacy in identifying the signature of assembly bias and Modified Gravity (MG). In the investigation of assembly bias, utilizing galaxy clustering data ranging from scales of $0.1 \sim 60 \Mpch$, we discover that our emulator model accurately recreates the cosmological parameters, $\Omega_m$ and $\sigma_8$, along with the velocity bias $\gamma_{f}$, staying well within the 1-$\sigma$ error margins, provided that assembly bias is considered. Ignoring assembly bias can considerably alter our model constraints on parameters $\Omega_m$ and $\sigma_8$ if the test sample includes assembly bias.
Using our emulator for MG simulations, which encompasses two Dvali-Gabadadze-Porrati models (DGP; N1, N5) and two $f(R)$ models (F4, F6), we can effectively identify a robust signature of modified gravity, for models such as DGP-N1 and $f(R)$-F4, as indicated by a noticeable deviation of $\gamma_{f}$ from unity. 
Using the velocity dispersion of dark matter halos to effectively represent the underlying strength of the velocity field of these MG simulations, we find that the simple $\gamma_{f}$ model can recover the truth with minimal bias. 
These evaluations indicate that our simple halo-velocity bias model is capable of detecting significant MG characteristics, although additional methodologies should be pursued to improve model constraints.
\end{abstract}


\section{Introduction}
\label{sec:Introduction}

Clustering measurement is one of the crucial outputs from large-scale galaxy surveys. The analyses at both large and small scale have provided important results for cosmology and galaxy formation physics as demonstrated by multiple survey programms such as the Sloan Digital Sky Survey (SDSS-I/II, \citealt{SDSS_York, Abazajian_2009}), the Two degree Field Galaxy Redshift Survey (2dFGRS, \citealt{Colless_2001, Cole_2005}), WiggleZ (\citealt{Drinkwater_2010}), SDSS-BOSS (\citealt{Dawson_BOSS}), SDSS-eBOSS (\citealt{eBOSS_Dawson}),  the Dark Energy Spectroscopic Instrument (DESI, \citealt{DESI_2016, DESI_2024_BAO}) and so on. With the large volume and high number density of different types of tracers, we can accurately measure the clustering statistics covering a wide spatial scale. This requires us to develop a sufficient model and technique to fully retrieve the underlying information from these data, especially at small scale since the measurement is sensitive to both fundamental cosmological model and galaxy formation physics.

In order to do so, \citet{Zhai2019, Zhai2023a} developed an emulator approach to make use of small scale clustering measurements for constraining cosmological parameters, based on the Aemulus Project (\citealt{DeRose2019}) which consists multiple N-body simulations aiming at precision emulation of galaxy survey observables. The model requires three components, N-body simulations with different cosmological parameter to describe the non-linear dynamics of dark matter density field, empirical and parameterized halo occupation distribution (HOD) model to connection galaxies and dark matter halos, and a Gaussian Process (GP) algorithm to enable fast interpolation in the parameter space. Due to the nontrivial coupling between cosmological parameters and galaxy-halo connection models, the interpretation of the result needs careful examination as we have done in the follow-up works from both the observational data side and modeling side (\citealt{Zhai_2023b, Zhai_2024}). On the other hand, it implies that the power of clustering analysis at non-linear scale is beyond the constraint on the cosmological parameters \citep{Yuan2022,Lange2022,Wibking2019,Salcedo2022,Storey2024}. In this work, we move one step further in this direction and investigate applications of this model. In particular, we examine the impact of assembly bias in cosmological inference and the possibility of measuring the signals of modified gravity models.

Assembly bias refers to the phenomenon where the clustering properties of dark matter halos depend not only on their mass but also on additional factors, such as their formation history or environment \citep{Sheth2004, Gao2005, Wechsler2006, Hearin2016, Villarreal2017, Wechsler2018, Mao2018, Shi2018, Contreras2023, Cuesta2023}. If the galaxy content in the dark matter halos is correlated with the halo assembly history, the halo occupation can vary with similar secondary halo properties and thus yield galaxy assembly bias (\citealt{Xu2021}). Understanding and quantifying assembly bias is crucial to accurately modeling and predicting the distribution of cosmic structures. 
Our previous work has already implemented an environment-dependent model for assembly bias, and the result shows that the significance is not strong in the BOSS galaxies (\citealt{Zhai2023b}). However, there are debates about the detection of this signal in the literature \citep{Salcedo2018, Shi2018, Zentner2019}. For instance, a density-split analysis of the same galaxy sample gives a preference of assembly bias of a few $\sigma$ \citep{Paillas2024}. In this work, we do not aim at examining the robustness of these measurements but instead focus on the modeling side and investigate the impact of assembly bias on the predictions of our model. Specifically, we examine the correlations between the deviations of the cosmic structure growth rate, $f\sigma_{8}$ and the strength of assembly bias signal arisen from clustering measurement. We hypothesize that incorporating relevant parameters associated with assembly bias into the emulator model is necessary to achieve more accurate and realistic predictions.

The other model we investigate in this work is the Modified Gravity (MG) theory. This type of model proposes modifications to the fundamental laws of gravity described by GR by introducing additional parameters or alternative equations of motion. Modified gravity models can address certain unresolved issues in cosmology, such as the accelerated expansion of the universe or the nature of dark matter, providing new avenues for understanding cosmic phenomena (\citealt{Clifton_2012}). In this paper, we consider four different methods for modifying gravity, including two models based on the Dvali-Gabadadze-Porrati (DGP) framework \citep{Dvali2000}, and two models based on $f(R)$ gravity \citep{Barraco1999, Capozziello2005, Felice2010, Sotiriou2010}. It is worthwhile to investigate their dynamic properties at different scales. By using an emulator model to focus on galaxy clustering at highly non-linear scales, we can investigate the viability and performance of these modified gravity models. We choose a method without computational demands of running numerous full N-body simulations but can verify the behavior of these modified gravity approaches. Given the empirical model we have developed, it also enables the search for MG effect at small scale, which is different compared to many current works \citep{Jain2008,Beutler2012,Desjacques2018,Frusciante2020} that focus on observables at linear scales. We anticipate that the analysis can offer alternative perspectives on the nature of gravity and its effects on the large-scale structure of the universe \citep{Samushia2014, Karcher2024, Ruan2022}.

The structure of this article is organized as follows. In Section \ref{sec:Simulation}, we provide a comprehensive description for the simulations used in this work, along with  the theoretical background of our model. Section \ref{sec:Analysis method} elaborates on the analysis method, including the two point correlation function (2PCF) and the likelihood analysis. Section \ref{sec:Results} describes the results, and finally, we summarize and conclude in Section \ref{sec:Conclusion}.

\section{Simulation}
\label{sec:Simulation}

In this section, we introduce the simulations used in this paper, including the Aemulus suite for building and testing the emulator of galaxy clustering and the modified gravity simulation to investigate the small-scale behavior. 

\subsection{The Aemulus suite}
\label{subsec:Aemulus}

The Aemulus suite from \citet{DeRose2019} includes 40 simulations of different cosmological parameters with independent initial phases to build emulators for statistics of dark matter halos and galaxies, as well as another 35 simulations to evaluate the emulator performance. All simulations are performed assuming General Relativity and have a box size of $(1.05 \Gpch)^{3}$ that contains $1400^{3}$ particles, resulting in a mass resolution of $3.51 \times 10^{10} {\Omega_{m}/0.3} \Msunh$. The dark matter halos are identified with the Rockstar algorithm \citep{Behroozi2013} and the resolution is able to model massive galaxies as explored in this paper. For more details on the simulation suite, we refer the reader to \citet{DeRose2019}. In this work, we focus on the galaxy sample with a number density and redshift that can match our earlier work in \citet{Zhai2023b} for the analysis of BOSS galaxies.

\subsection{MG simulations}
\label{subsec:MG Simulations}

In this subsection, we describe the modified gravity simulations adopted in this work, the chameleon $f(R)$ gravity and the DGP braneworld models.

\subsubsection{$f(R)$ gravity}
\label{ssubsec:f(R)}

The $f(R)$ gravity \citep{Sotiriou2010} expands upon general relativity (GR) by modifying the Einstein-Hilbert action to include a generic function $f(R)$ dependent on the Ricci scalar $R$, thereby permitting a more versatile depiction of gravitational forces. Consequently, the form of the Einstein-Hilbert action is given by \citep{Carroll2001}
\begin{equation} \label{eq:S}
S=\int \mathrm{d}^4 x \sqrt{-g}\left[\frac{R+f(R)}{16 \pi G}+\mathcal{L}_{\mathrm{M}}\right]\,.
\end{equation}
The metric tensor is indicated by $g_{\mu \nu}$, and $g$ signifies the determinant of $g_{\mu \nu}$. $G$ is the universal gravitational constant, and $\mathcal{L}_{\mathrm{M}}$ is the Lagrangian density for the matter field. The Ricci scalar is given by $R=12 H^2+6 H H^{\prime}$. 
The corresponding field equation can be written as
\begin{equation} \label{eq:G}
G_{\mu \nu}+f_R R_{\mu \nu}-\left(\frac{1}{2} f-\square f_R\right) g_{\mu \nu}-\nabla_\mu \nabla_\nu f_R=8 \pi G T_{\mu \nu}^m\,.
\end{equation}

The behavior of the $f(R)$ theory depends on the specific form of the chosen function $f(R)$ \citep{Sahlua2024, Saadeh2024, Vogt2024,  Hu2007}. Different choices of $f(R)$ can lead to different cosmological predictions and gravitational dynamics. 
In this paper, we employ the $f(R)$ function derived from the Hu-Sawicki model \citep{Hu2007}, which can produce a phase of late-time accelerated expansion of the universe
\begin{equation} \label{eq:f(R)}
f(R)=-m^2 \frac{c_1\left(-R / m^2\right)^n}{c_2\left(-R / m^2\right)^n+1}
\end{equation}
where $m^2 \equiv 8 \pi G \bar{\rho}_{\mathrm{M}, 0} / 3=H_0^2 \Omega_{\mathrm{m0}}$, $\bar{\rho}_{\mathrm{M}, 0}$ is the background matter
density at $z = 0$, $H_0$ is the Hubble constant and $\Omega_{\mathrm{m0}}$ the dimensionless matter density parameter at today. $c_1$; $c_2$ and $n$ are free model parameters. The parameter $n$ is a positive number, which is set to $n = 1$ in most previous studies of this model \citep{Leizerovich2022,Fakhry2024,Yan2024,Aviles2025}
There is a limit where $f(R) \approx -m^2c_1/c_2$ when $R \gg m^2$, which corresponds to the well-known Einstein-Hilbert action with cosmological constant. 

\subsubsection{DGP model}
\label{ssubsec:DGP}

Another modified gravity method is the Dvali-Gabadadze-Poratti model \citep[nDGP model]{Dvali2000}. The idea of the theory is the assumption that our universe is a four-dimensional membrane embedded in a five-dimensional spacetime (bulk), which is often referred to as the cosmic membrane model. There is a self-acceleration branch of solution (sDGP), which gives a natural explanation for the cosmic acceleration but is inconsistent with observations such as the CMB, supernovae, and local measurements of $H_0$ \citep{Song2007,Fang2008,Capilla2020,Ruan2024},
and a normal branch DGP (nDGP) model \citep{Koyama2007} which introduces an additional component (gravity spreading in higher dimensions) rather than cosmological constant driving the accelerated expansion of the universe. In this work, we only consider the normal branch solution.
The gravitational action of the model is given by 
\begin{equation} \label{}
\begin{split}
    S=& \int_{\text {brane }} \mathrm{d}^4 x \sqrt{-g}\left(\frac{R}{16 \pi G}\right) \\
    & +\int_{\text {bulk }} \mathrm{d}^5 x \sqrt{-g^{(5)}}\left(\frac{R^{(5)}}{16 \pi G^{(5)}}\right)
\end{split}
\end{equation}
where $G^{(5)}$ is one new parameter in nDGP model which denotes the quantity in the five-dimensional bulk. And it can be obtained from the crossover scale $r_c \equiv G^{(5)}/{2 G}$, above which gravity starts to have a non-standard 5-dimensional behavior \citep{Maartens2010}.
The modified Friedmann equation in a homogeneous and isotropic universe is 
\begin{equation}
    \frac{H(a)}{H_0}=\sqrt{\Omega_{m 0} a^{-3}+\Omega_{\mathrm{DE}}(a)+\Omega_{\mathrm{rc}}}-\sqrt{\Omega_{\mathrm{rc}}}
\end{equation}
where $\Omega_{\mathrm{rc}} \equiv 1/(4 H^2_0 r^2_c)$. The combination of parameters $H_0 r_c$ is used to quantify the deviations from GR. The Friedmann equation returns to the $\Lambda$CDM case where $H_0 r_c \rightarrow \infty$. The deviation from GR will be stronger when the value of $H_0 r_c$ is smaller. 

The cosmological structure formation in nDGP model is governed by the modified Poisson and scalar field equations \citep{Koyama2007}:
\begin{equation} \label{eq:Phi}
\nabla^2 \Phi=4 \pi G a^2 \delta{\rho}_{\mathrm{m}} +\frac{1}{2} \nabla^2 \varphi
\end{equation}
\begin{equation} \label{eq:phi}
\nabla^2 \varphi+\frac{r_c^2}{3 \beta a^2 c^2}\left[\left(\nabla^2 \varphi\right)^2-\left(\nabla_i \nabla_j \varphi\right)^2\right]=\frac{8 \pi G a^2}{3 \beta} \delta \rho_{\mathrm{m}}
\end{equation}
where $\varphi$ is a new scalar degree of freedom, $\delta\rho_{\mathrm{m}} = \rho_{\mathrm{m}} - \bar{\rho}_{\mathrm{m}}$.
Here, the scalar field, or more precisely its spatial perturbation, functions as the potential for this supplementary force. The efficacy of these screening mechanisms is intricately linked to the dynamics of the scalar field, which is dictated by its equation of motion.

\subsubsection{Simulations}
\label{ssubsec:Simulations}

Based on these two classes of modified gravity models, we use the ELEPHANT (Extended LEnsing PHysics using ANalytic ray tracing) suite of MG N-body simulations to enable our analysis \citep{Cautun2018, Paillas2019, Aguayo2019}.
These simulations are run from redshift 49 to 0 with a box size of 1024$\Mpch$ including $1024^3$ DM particles with $m_p = 7.798 \times 10^{10}h^{-1}M_{\odot}$. The same background cosmological parameters are applied: $\Omega_m=0.281$, $\Omega_b=0.046$, $\Omega_{\mathrm{CDM}}=0.235$, $\Omega_\Lambda=0.719$, $\Omega_\nu=0$, $h=0.697$, $n_s=0.971$, $\sigma_8=0.820$. $f(R)$ and DGP modified gravity simulations are run using the ECOSMOG and ECOMOG-V code, respectively \citep{Li2012,Li2013}. The ELEPHANT simulation suite has a fiducial $\Lambda$CDM model, 3 variants of $f(R)$ gravity with $n=1$ and $f_{R0}=-1\times10^{-4,-5,-6}$ called F4, F5 and F6, and 2 variants of nDGP model with $H_0r_c=1,5$ called N1 and N5. Note that we do not present results for F5 $f(R)$ gravity, since its behavior is in the middle between F4 and F6. Details of these simulations are shown in Table \ref{tab:MG_models}.
Similarly to the Aemulus suites, dark matter halos are also identified by the ROCKSTAR halo finder, but with some changes due to the fifth force of modified gravity effect \citep{Gupta2024}.  
However, \citet{Li2010} found that the effect of modification is very small, so the standard ROCKSTAR algorithm is used on both the $\Lambda$CDM and MG models. With these halo catalogs from the MG simulations, we can produce galaxy mocks and investigate their properties compared with GR as described in the following sections.

\begin{table}[htbp]
    \centering
    \caption{Cosmological models with different modified gravity.}
    \begin{tabular}{p{1cm}|p{6cm}}
        \hline \hline
        Model & Description  \\
        \hline
        GR & $\Lambda$CDM w/o massive neutrinos in WMAP9 cosmology \\
        \hline
        N5 & nDGP model with $\Lambda$CDM background and $H_0r_c=5$ (closest to GR) \\
        N1 & nDGP model with $\Lambda$CDM background and $H_0r_c=1$ (stronger deviation from GR) \\
        \hline
        F6 & Hu-Sawicki $f(R)$ with $n=1$ and $f_{R0}=-1\times10^{-6}$ (closest to GR) \\
        F4 & Hu-Sawicki $f(R)$ with $n=1$ and $f_{R0}=-1\times10^{-4}$ (stronger deviation from GR) \\
        \hline \hline
    \end{tabular}
    \label{tab:MG_models}
\end{table}

\subsection{Generating Mock samples based on HOD model}
\label{subsec:HOD}

We use the HOD approach to model the connection between galaxies and dark-matter halos. With the Aemulus and MG simulations, we can directly produce galaxy mocks and measure the clustering signals. Then we focus on non-linear scale and apply the emulator model to investigate the impact on the cosmological measurement.
The fundamental representation in the basic HOD is the likelihood that a halo with mass M harbors N galaxies of a specified type, denoted by $P(N/M)$. This offers a depiction of how galaxies are spatially distributed within the dark-matter halo, assuming that the galaxy distribution relies solely on the halo's mass. The basic HOD parametrization is separated to the contribution of the central and satellite galaxies with the mean occupancy of halos:
\begin{equation}
    \langle N(M)\rangle=\left\langle N_{\mathrm{gal}}(M)\right\rangle=\left\langle N_{\mathrm{cen}}(M)\right\rangle+\left\langle N_{\mathrm{sat}}(M)\right\rangle
\end{equation} 
The functional form follows the early work of \citet{Zheng2005} but is slightly updated in \cite{Zhai2019}.
In total, there are five parameters in the basic HOD model, which includes $M_{\mathrm{min}}$, the mass at which half the halos have a central galaxy, $M_{\mathrm{sat}}$, the typical mass scale for halos to host one satellite, $\alpha$, the power law index for the mass dependence of the number of satellites, $M_{\mathrm{cut}}$, the mass cutoff scale for the satellite occupation function, and $\sigma_{\log M}$, the scatter of halo mass at fixed galaxy luminosity. Since we aim to model samples of the BOSS-CMASS galaxies in this work, we have an additional constraint on the galaxy number density $n_{gal}$. Therefore, we use $n_{gal}$ as another parameter and solve $M_{\mathrm{min}}$ when all other parameters are given.
In addition, we also consider the extended parameters including $\eta_{\mathrm{con}}$, concentration of satellite distribution relative to the dark-matter halo, $\eta_{\mathrm{vc}}$ and $\eta_{\mathrm{vs}}$, the velocity bias parameter for the central and satellite galaxies \citep{Guo2015}. 

In addition to the above parameter set, we also consider models for galaxy assembly bias. The absence of a shared understanding on observational limitations concerning galaxy assembly bias allows for a versatile selection of secondary halo characteristics to explore this bias, encompassing both intrinsic and extrinsic attributes. In this research, we focus on the bias caused by the environment, an external characteristic which is defined as the dark matter overdensity $\delta$ within $10 \Mpch$  of the halos \citep{McEwen2018}. 
Taking this into account, we adjust the HOD model by scaling the parameter $M_{\mathrm{min}}$ \citep{Zhai2023a}.
\begin{equation}
    \bar{M}_{\mathrm{min}}=M_{\mathrm{min}}\left[1+f_{\mathrm{env}} \operatorname{erf}\left(\frac{\delta-\delta_{\mathrm{env}}}{\sigma_{\mathrm{env}}}\right)\right]
\end{equation}
where the amplitude parameter $f_{\mathrm{env}}$ regulates the overall strength of the dependence and the resulting level of assembly bias, the position parameter $\delta_{\mathrm{env}}$ sets the threshold to separate halos residing in over- and under-dense regions, and the width parameter $\sigma_{\mathrm{env}}$ governs the smoothness of the transition from under-density to overdensity. 
This assembly bias model enables the minimum mass scale for dark-matter halos to host a central galaxy to be contingent on halo environment, which directly affects the occupancy of centrals and satellites and consequently can alter the clustering signal compared to the basic HOD model. 
Similarly to other properties that can be modeled for galaxy assembly bias, their impact on clustering measurement can propagate to the inference of the galaxy halo connection model and cosmological parameters (\citealt{Zentner2014}). The first part of this work is to explicitly investigate the possible bias induced by galaxy assembly bias when we perform cosmological analysis. We note that there are many studies in the literature to search for the signal and its significance; however, the results are not conclusive; see, for instance, \cite{Salcedo2018, Zentner2019} and references therein. Our work serves as an attempt from the theoretical point of view and we provide more details in Section \ref{subsec:AB}.

In addition to the above model components, we consider another phenomenological parameter $\gamma_{f}$, which is defined as 
\begin{equation}
    \gamma_{f} = f / f_{GR}\,,
\end{equation}
and used to scale the amplitude of halo velocity field relative to the prediction from our model of $w\mathrm{CDM}+\mathrm{GR}$.
This parameter was first introduced in \cite{Reid_2014} to measure $f\sigma_{8}$ based on a fixed cosmological model, since it is shown that a fractional change in $\gamma_{f}$ is approximately equal to a fractional change in the amplitude of the peculiar velocity field on a large scale and $f\sigma_{8}$. In the subsequent works of \cite{Zhai2019} and \cite{Zhai2023a}, we adopt this degree of freedom to increase the parameter space range determined by the priors of the simulation suite to enable a more robust constraint on $f\sigma_{8}$. This model is further generalized in \cite{Chapman2023} to split the velocity scaling into a linear and non-linear part for the measurement of $f\sigma_{8}$. It makes the model more flexible and accurate with an additional parameter, but for our purpose, the single parameter model is sufficient. In addition, the $\gamma_{f}$ parameter can also mimic the behavior of certain types of modified gravity if the impact is to change the amplitude of the velocity field. In this work, we explicitly investigate this property with multiple modified gravity simulations and present the results and details in Section \ref{subsec:MG}.

\section{Analysis method}
\label{sec:Analysis method}

\subsection{Galaxy clustering statistics}
\label{subsec:Clustering}
 
We use the two-point correlation function (2PCF) $\xi(r)$ to characterize the clustering signal, which is defined as the measurement of the excess probability of finding two galaxies separated by a vector distance $\mathbf{r}$, relative to a random distribution.  
To minimize the impact of redshift-space distortions and obtain information in real space (denoted by the subscript R in the subsequent equation), we calculate the projected correlation function \citep{Davis1983} with 
\begin{eqnarray}
    w_p\left(r_p\right) &=& 2 \int_0^{\infty} d \pi \xi_Z\left(r_p, \pi\right) \nonumber \\
   &=& 2 \int_0^{\infty} d \pi \xi_R\left(r=\sqrt{r_p^2+\pi^2}\right)\,.
\end{eqnarray}
The integrand is truncated to $\pi_{\mathrm{max}}=80\Mpch$, which is believed to be large enough to contain most of the correlated pairs and give stable results. In addition, we use the standard decomposition with Legendre polynomial to obtain the multiples of correlation function
\begin{equation}
    \xi_{l}(s) = \frac{2l+1}{2} \int_{-1}^{1} L_l(\mu)\xi_Z(s,\mu) d\mu \,,
\end{equation}
where $\mu=r_p /s$, $L_l$ is the Legendre polynomial of order $l$. In the analysis, we adopt $\xi_0$ and $\xi_2$ since most of the information in the redshift space is contained in the first few multiples.

For the galaxy samples used in this work, we estimate the two point correlation function through the Landy-Szalay estimator \citep{Landy1993}
\begin{equation}
    \hat{\xi}(r) = \frac{DD-2DR+RR}{RR}\,,
\end{equation}
where DD, DR and RR are data-data, data-random, and random-random pair counts normalized by the total number of possible pairs in a given radial bin. Following our earlier works, we choose logarithmically spaced bins for $r_p$ or $s$ from the scale $0.1\sim60\Mpch$ in the calculation of the correlation function, resulting in nine data points for each statistic. 

\subsection{Likelihood analysis}
\label{subsec:Likelihood}

In this work, we investigate the impact of both galaxy assembly bias and modified gravity on the constraint on model parameters. This is quantified through a Bayesian analysis with likelihood function
\begin{equation} \label{eq:likelyhood}
\ln \mathcal{L} = -\frac{1}{2}(\xi_{\rm emu} - \xi_{\rm obs}) C^{-1} (\xi_{\rm emu} - \xi_{\rm obs})
\end{equation}
where $\xi_{\mathrm{emu}}$ and $\xi_{\mathrm{obs}}$ are the correlation function from our model and observational data, respectively, and $C$ is the covariance matrix. Depending on the tests, $\xi_{\mathrm{obs}}$ is a data vector from different galaxy mocks, i.e. produced by different galaxy assembly parameters, or using modified gravity simulations. We adopt the parallelized MULTINEST package \footnote{\href{https://github.com/JohannesBuchner/MultiNest}{https://github.com/JohannesBuchner/MultiNest}} \citep{Feroz2009,Buchner2014} to explore the parameter space.
The code is based on the nested sampling algorithm \citep{Skilling2004}. Its output can provide Bayesian evidence and posterior distributions of the parameters. 
In our analysis, we chose 1000 livepoints to sample the high-dimensional parameter space until the convergence criteria for Bayesian evidence are reached.

The other component of our analysis is the covariance matrix. 
Similar to our previous works in \citep{Zhai2023b} to match a galaxy sample from the BOSS survey in terms of the number density and redshift, the covariance matrix in our analysis comes from two sources expressed as:
\begin{equation} \label{eq:cov_matrix}
C = C_{\rm sam} + C_{\rm emu}\,,
\end{equation}
where $C_{\rm sam}$ stands for sample variance estimated from jackknife resampling as in \citet{Zhai2023a}. 
Note that we used the fractional error from this BOSS-CMASS analysis and scaled the uncertainty to match the corresponding volumes. We also adopt some smoothing algorithm in the correlation matrix to avoid spurious results due to noisy eigen vectors, but the earlier analysis shows that this does not have a significant impact (\citealt{Zhai2023a}). The second component $C_{\rm emu}$ corresponds to the intrinsic error of the emulator, which can be calculated using the same method as in \citet{Zhai2019} simply by comparing the predictions of our emulator with the test simulations.

\begin{figure*}
    \centering
    \includegraphics[width=0.45\linewidth]{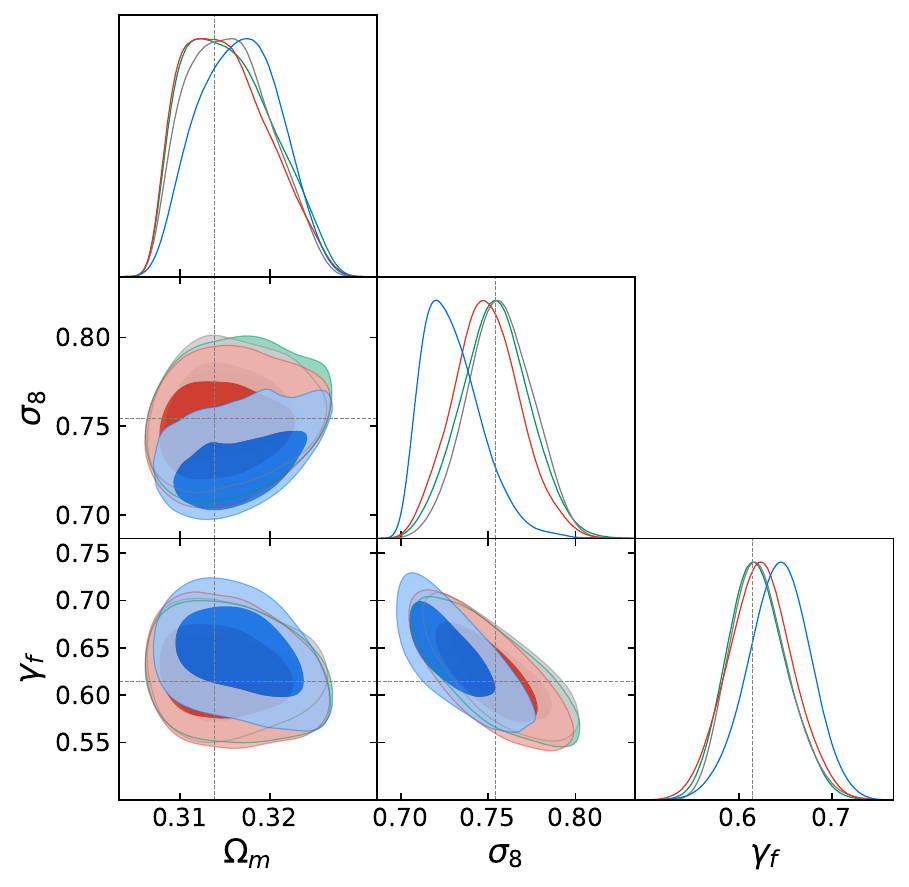}
    \includegraphics[width=0.45\linewidth]{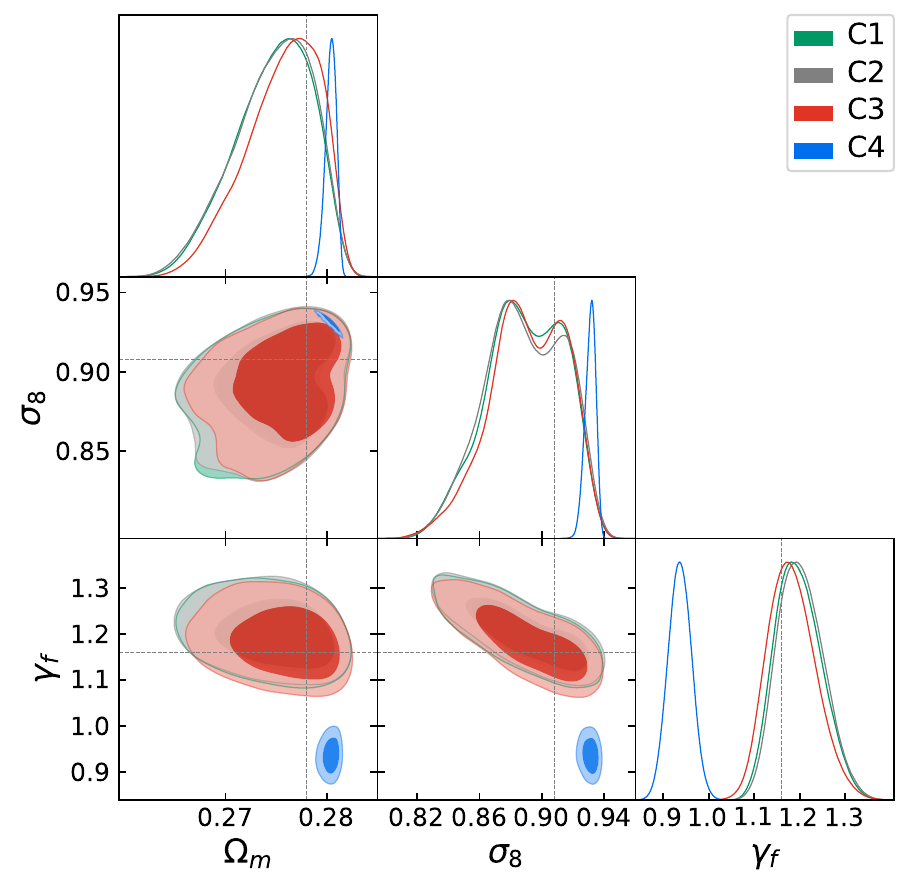}
    \caption{Two examples of the constraints on $\Omega_{m}$, $\sigma_8$ and $\gamma_{f}$ for two different cosmology and different HOD models. The `C1, C2, C3, C4' for different colors stand for different cases which are listed in Table \ref{tab:sim_emu}. The first three fit perfectly but there is the largest deviation for `C4' within blue regions, indicating that ignoring assembly bias can yield biased constraints on cosmological parameters. The deviation between cases with and without assembly bias in these two panels can be quantified by $\Delta\xi_i/\xi_i (s=10~h^{-1}\mathrm{Mpc})$ as explained in the text. Specifically, the deviations are -0.0941 for the left panel and -0.3577 for the right panel, i.e. the strength of assembly is roughly proportional to the bias in the final constraints. }
    \label{fig:AB_contour}
\end{figure*}

\section{Results}
\label{sec:Results}

In this section, we present the result of applying our emulator model to various simulations.
For GR simulations, we study the importance of assembly bias in constraining cosmological parameters. 
In MG simulations, we focus primarily on investigating the effectiveness of identifying the modified gravity signal and constraining parameters. 

\subsection{Application of the Aemulus suite to test galaxy assembly bias}
\label{subsec:AB}

\begin{figure}[htb]
    \centering
    \includegraphics[width=0.48\textwidth]{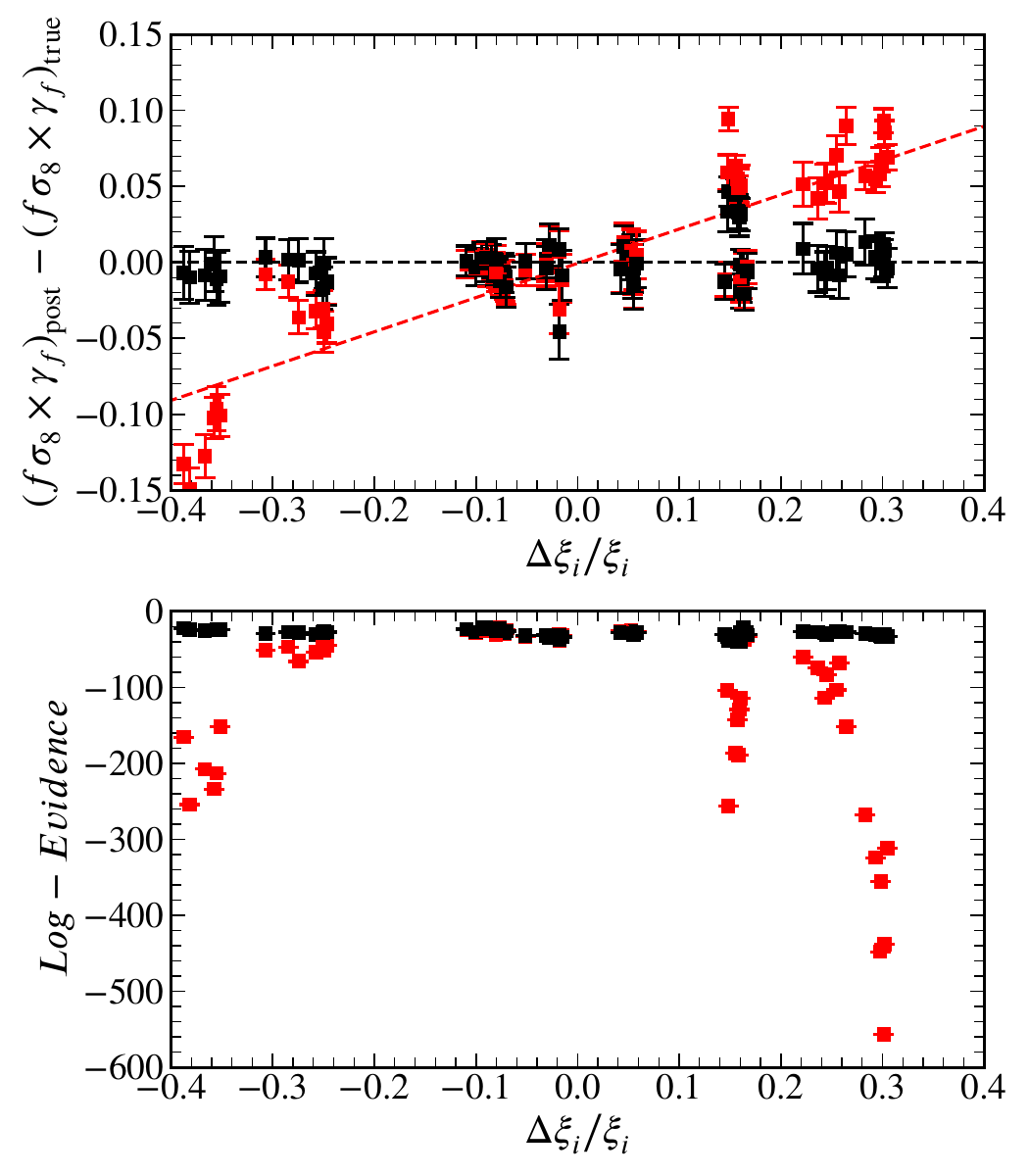}
    \caption{Bias in the measurement of $f\sigma_{8}$ when the model of galaxy clustering at non-linear scale has assembly bias or not. The mocks are produced by the Aemulus test simulations with different HOD models and assembly bias parameters. In order to denote the strength of assembly bias, for each HOD model we compute the residual of $\xi_{i}$ compared with the result when $f_{\mathrm{env}}$ is set to 0, i.e. the assembly bias effect is turned off. We arbitrarily choose the value at $10\Mpch$ as a proxy and plot as the x-axis, which is equivalent to the y-axis in Figure \ref{fig:AB_10MPC}. Then for the test model we run recovery tests with (black squares) and without (red squares) assembly bias parameters in the model. The result shows that the ignorance of assembly bias in the model can lead to biased cosmological measurement when the universe has assembly bias in it, and the deviation is proportional to the strength of the assembly bias. The upper panel shows the offset of $f\sigma_{8}$ as the signal, while the bottom panel shows the Bayesian evidence for the corresponding tests. It shows that a incomplete model for assembly bias is not favored by the data, consistent with the biased measurement in the upper panel. }
    \label{fig:AB_deviation}
\end{figure}

We primarily evaluate the impact of galaxy assembly bias on cosmological inference, using two sets of mock galaxy samples:
one that includes assembly bias and one that does not. These mocks are generated by randomly selecting 10 HOD models from the test samples which are originally used to evaluate the emulator performance as discussed in Section \ref{subsec:Aemulus}, applied across all test simulations with 7 different configurations of cosmological parameters. This results in $70\times 2$ sets of mocks. Although these HOD models are randomly picked, we require that their clustering amplitude at large scale not radically offset compared to BOSS-CMASS. The appearance of assembly bias in these test models is simply determined by the parameter $f_{\mathrm{env}}$ in the production and $f_{\mathrm{env}}=0$ means the mock has no environment-dependent assembly bias. Based on this set of mocks with and without assembly bias, we investigate the constraining ability of the emulator in the aspect of galaxy distribution.
Using these mock galaxy samples, we create four scenarios, each differing in whether assembly bias is present in the test mocks and whether the model settings include assembly bias parameters, as outlined in Table \ref{tab:sim_emu}. We explore $70 \times 4$ different parameter configurations in total. 

\begin{table}[htbp]
    \centering
    \caption{Four combinations of test mocks and emulator model with/without assembly bias corresponding to Figure \ref{fig:AB_contour}.}
    \begin{tabular}{c|c|c}
        \hline \hline
         & Mock galaxy sample & Model \\
        \hline
        C1 & without assembly bias & with assembly bias \\
        C2 & without assembly bias & without assembly bias \\
        C3 & with assembly bias & with assembly bias \\
        C4 & with assembly bias & without assembly bias \\
        \hline \hline
    \end{tabular}
    \label{tab:sim_emu}
\end{table}

Figure \ref{fig:AB_contour} shows two examples of the constraints on a subset of the key cosmological parameters $\Omega_{m}$, $\sigma_8$ and $\gamma_{f}$ of two models for illustration purpose. There are four colors in each panel that correspond to four different scenarios, as summarized in Table \ref{tab:sim_emu}.
Our findings suggest that the first three scenarios align closely with the parameter inputs for the HOD models. This is consistent with our expectations and ensures that the emulator is constructed without bias. However, in the fourth scenario shown in blue, where the mock galaxy sample mirrors the real universe and includes assembly bias, but the emulator model does not activate related parameters ($f_{\mathrm{env}}, \delta_{\mathrm{env}}, \sigma_{\mathrm{env}}$)—a significant discrepancy emerges for different parameters, and it can be as large $5\sigma$ for $\gamma_{f}$ within certain HOD models. This outcome supports our hypothesis: incorporating assembly bias parameters in the emulator model is crucial.
Recent studies have increasingly highlighted the link between galaxy distribution and secondary properties within dark-matter halos \citep{Wechsler2018, Mao2018, Shi2018, Contreras2023}, and the galaxy's environment can impact its distribution and clustering properties (\citealt{McEwen2018}). 
\citet{Cuesta2023} found that ignoring assembly bias can bias $\sigma_{8}$ downward, especially with small-scale data for SDSS LOWZ-like samples.
Consequently, integrating assembly bias effects into future universe predictions is essential. Our emulator model provides robust confidence in understanding how assembly bias influences cosmological parameters.

In addition, we observe that the deviation in the fourth case due to different HOD parameters is not identical. To delve deeper into how different HOD models influence parameter constraints, we adopt the fraction of the clustering deviation between measurements with and without assembly bias to assess the impact of assembly bias of the HOD mocks. We present this effect in Figure \ref{fig:AB_10MPC} of the appendix, particularly at $10 \Mpch$, with varying color intensities representing different HOD models. This leads us to examine the link between assembly bias-induced clustering bias across HOD models and the cosmological parameter deviations constrained by the emulator model. 

To simplify the comparison, we focus on $f\sigma_{8}$ to represent the cosmological measurements. We first compute the ratio of $\xi_{0}$ for the same HOD model with and without assembly bias effect and choose the residual at $10\Mpch$ as a proxy (see for instance Figure \ref{fig:AB_10MPC} in the appendix) to represent the strength of the assembly bias in terms of galaxy clustering. Note that the choice of this particular scale and summary statistics is not unique, and we find that applying other scales or statistics does not change our conclusion significantly. Using all these test mocks with assembly bias, we perform recovery tests and present the results as black and red squares, respectively, in the upper panel of Figure \ref{fig:AB_deviation}. It shows the bias in the cosmological measurement when the assembly bias is not properly taken into account in the model, and the offset is proportional to the strength of the assembly bias. Given the uncertainty on $f\sigma_{8}$ measurement of a CMASS-like survey, the offset for certain models of non-linear clustering can be as large as a few $\sigma$. Thus we can see that this effect does not only impact the construction of the correct galaxy-halo connection model as seen in the literature (e.g. \citealt{Zentner2014}), but can also influence the cosmological measurements. For the examples in Figure \ref{fig:AB_contour}, the ratio of $\xi_{0}(s=10^{-1}\mathrm{Mpc})$ between models with and without assembly bias is -0.09 and -0.35 respectively, consistent with our expectation that the bias in the parameter constraint is roughly proportional to the strength of assembly bias.

The lower panel shows the Bayesian evidence produced from our nested sampling analysis. The black squares with error bars depict models with assembly bias parameters, consistently matching the true values and yielding reasonable goodness of fit; therefore, our emulator model is capable of effectively constraining the cosmological parameters. In contrast, the red squares signify scenarios without assembly bias parameters in the model. The increased deviation on the X-axis intensifies the parameter limitation impacts, suggesting that a stronger effect of assembly bias on galaxy clustering enhances the constraint effect of related parameters in the model. Therefore, we conclude that the inclusion of the assembly bias parameter in the model is highly important for precisely constraining cosmological parameters.

\subsection{Application on the MG simulations}
\label{subsec:MG}

Our previous section is a simple and explicit test of the impact of galaxy assembly bias on the cosmological measurements within the Aemulus simulation based on GR+$w$CDM. In this section, we generalize to an external simulation suite and test the impact of MG, using similar methodologies. The ELEPHANT suite has a fiducial simulation of GR as a reference, and multiple modified gravity methods. We consider two DGPs (N5, N1) and two $f(R)$ (F6, F4) models in our analysis, as described in Table \ref{tab:MG_models}. 
Similarly to the above analysis of assembly bias, we also select 10 HOD models for each variant of modified gravity, all with identical cosmological parameters for the background evolution (further details are provided in Section \ref{ssubsec:Simulations} and Section \ref{subsec:HOD}). 

\begin{figure*}
    \centering
    \includegraphics[width=0.45\linewidth]{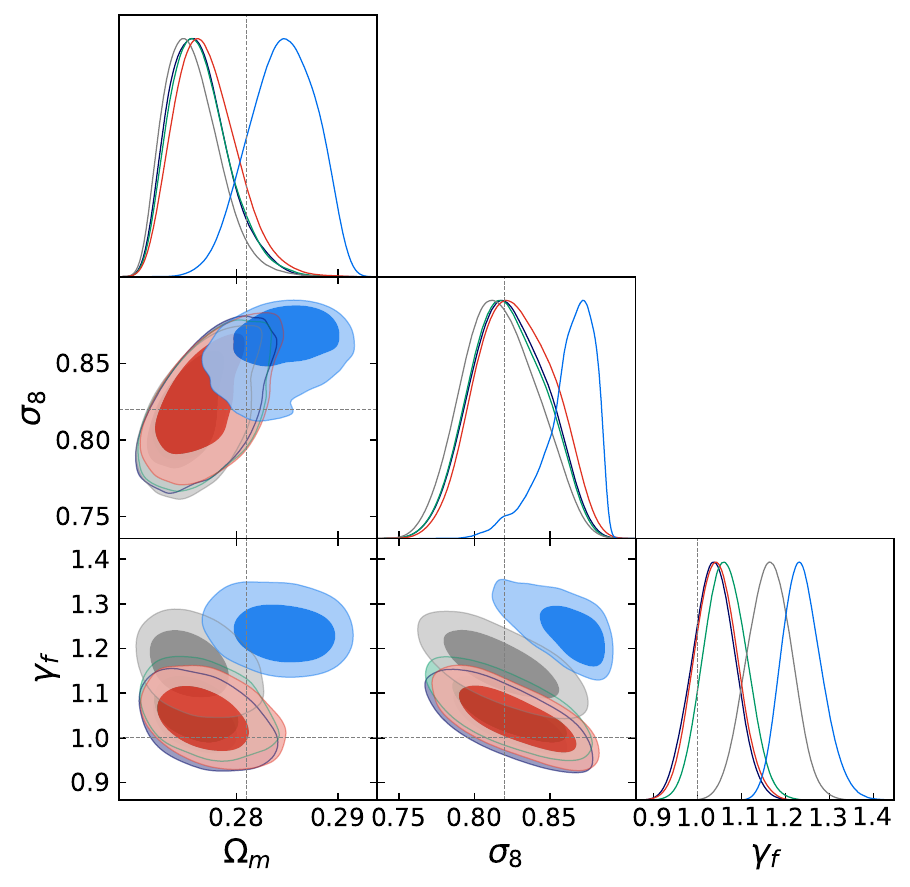}
    \includegraphics[width=0.45\linewidth]{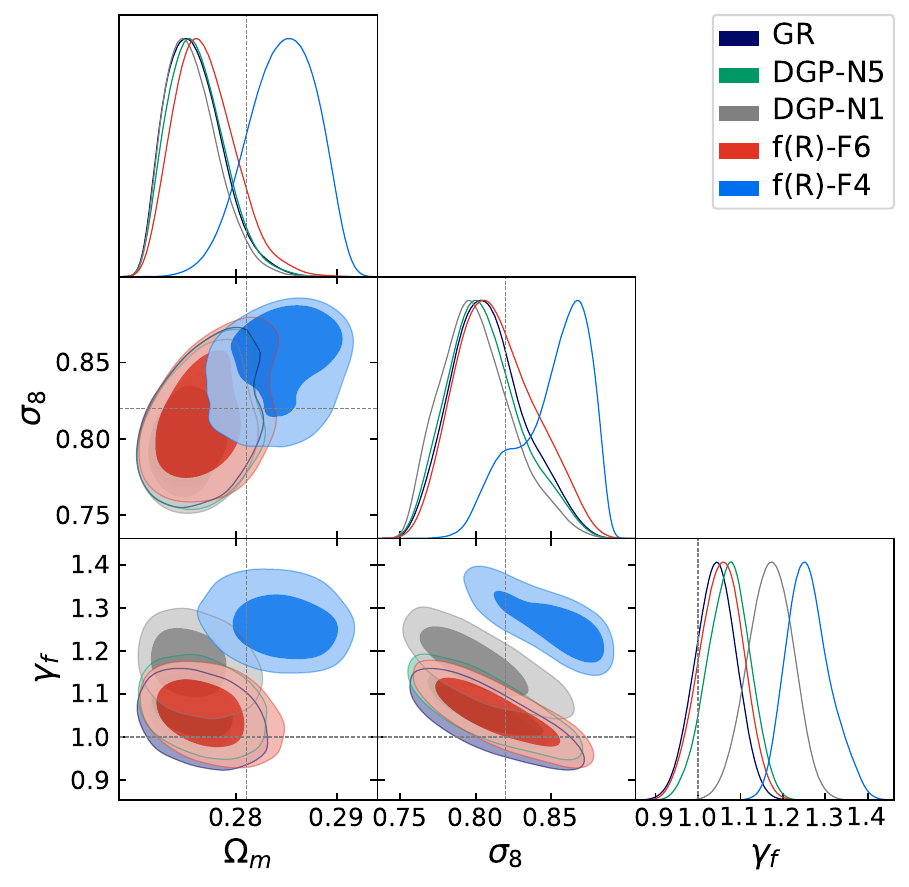}
    \caption{Two examples of the constraints on $\Omega_{m}$, $\sigma_8$ and $\gamma_{f}$ for two different HOD models using MG simulations. Different colors in each panel stand for different modified gravity methods which are listed in Table \ref{tab:MG_models}. Here we focus on the deviation of $\gamma_{f}$ from 1 which can approximately characterizes the intensity of the modified gravitational effect compared with GR，as well as constraints and degeneracies with other parameters. }
    \label{fig:MG-contour}
\end{figure*}

\begin{figure*}[!htb]
    \centering
    \includegraphics[width=1\textwidth]{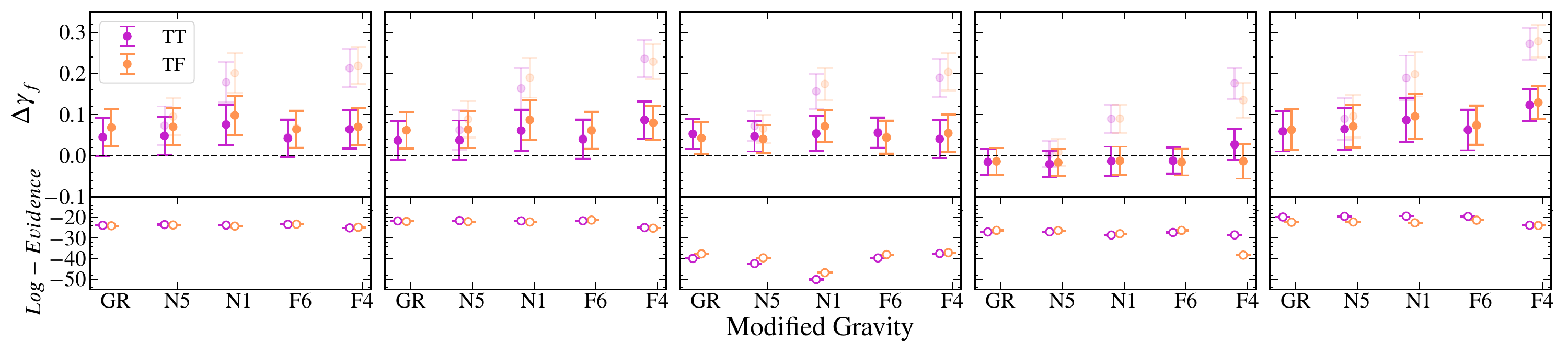}
    \caption{Constraints on $\gamma_{f}$ (upper half of each panel) from unity and Bayesian Evidence (lower half of each panel) for GR and four modified gravity methods with the same cosmological parameter setting at $z=0.55$. Five panels correspond to five different HOD models. The errorbars denote 1-$\sigma$ distributions. The meaning of cases `TT and TF' with different color are shown in Table \ref{tab:emu_MG}. It's worth noting that the true underlying velocity field is different in these MG models. We isolate an effective true $\gamma_{f}$ value of these MG simulations using the VD correction as explained in the text. The thick and thin symbols show the raw constraint and VD corrected results respectively.}
    \label{fig:MG}
\end{figure*}

With these test mocks, we adopt our emulator for galaxy correlation function based on Aemulus simulation and perform recovery tests on the key cosmological parameters as shown in Figure \ref{fig:MG-contour} and \ref{fig:MG}. Figure \ref{fig:MG-contour} presents two examples for constraints on a subset of the parameters, including $\Omega_{m}$, $\sigma_8$, and $\gamma_{f}$ for two HOD models that are randomly selected. Each panel within the figure encompasses five distinct colors. These colors respectively correspond to GR and four different modified gravity models. Due to the simulation specifics, such as the box volume, mass resolution, halo finder algorithms, snapshot redshift for the analysis, and so on, we do not expect perfect consistency and recovery for the reference GR simulation, but the result can serve as a reference, and we do expect this model to have a minimum bias compared with the MG models. 

In addition to the assembly bias effect addressed in the previous subsection, this study also aims to identify the deviation of $\gamma_{f}$ from unity as an indicator of the MG signature. We note that the modified gravity can have different impact on the statistics of both dark matter halo and galaxies, for instance the halo mass function or the scale dependent impact on the power spectrum (\citealt{Arnold_2019, Hernandez-aguayo_2019, Cataneo_2019}). Our clustering analysis at small scale assumes that the information extraction is dominated by the RSD effect due to peculiar velocity. Therefore we use the $\gamma_{f}$ model as a diagnostic to probe the underlying velocity field predicted by modified gravity models.
Figure \ref{fig:MG-contour} illustrates that within the framework of GR, the $\gamma_{f}$ value is approximately 1, aligning with expectations. For the other four modified gravity methodologies, the deviation of $\gamma_{f}$ from 1 increases in agreement with the increasing strength of the modified gravity. Consequently, the simplistic halo velocity bias model can serve as an effective and straightforward tracer, enabling the determination of salient characteristics of Modified Gravity. On the other hand, we find that the constraint on $\Omega_{m}$ is quite tight for all tests. Part of the reason is likely due to the fact that we keep the other cosmological parameters fixed at the true value and focus on the parameters that are the most sensitive to the growth rate measurement. This makes the constraint slightly offset from the true value of $\Omega_{m}$, but the significance for our test models is not higher than $2\sigma$. This can be attributed to the circumstance that distinct samples tend to produce appreciable systematic errors and sample variances with different HOD parameter configurations.

\begin{table}[htbp]
    \centering
    \caption{Four combinations of our model with/without velocity scaling parameter ($\gamma_{f}$) and assembly bias parameters ($f_{\mathrm{env}}$, $\delta_{\mathrm{env}}$, $\sigma_{\mathrm{env}}$) corresponding to the tests in Figure \ref{fig:MG}. }
    \begin{tabular}{l|c|c}
        \hline \hline
         & $\gamma_{f}$ & AB ($f_{\mathrm{env}}$, $\delta_{\mathrm{env}}$, $\sigma_{\mathrm{env}}$) \\
        \hline
        TT  & Turn on (T) & Turn on (T) \\
        TF  & Turn on (T) & Turn off (F)\\
        \hline \hline
    \end{tabular}
    \label{tab:emu_MG}
\end{table}

Due to the flexibility of our HOD model, we can investigate the correlations between the MG effect and the components of our galaxy-halo connection models such as the spatial distribution and kinematics of both central and satellite galaxies. As an example, we focus on galaxy assembly bias parameters as discussed in the previous section. For each test mock, we can apply different combinations of the velocity scaling parameter ($\gamma_{f}$) and the assembly bias parameters ($f_{\mathrm{env}}, \delta_{\mathrm{env}}, \sigma_{\mathrm{env}}$) in our model as presented in Table \ref{tab:emu_MG} to obtain the final constraints.

\begin{figure}
    \centering
    \includegraphics[width=1\linewidth]{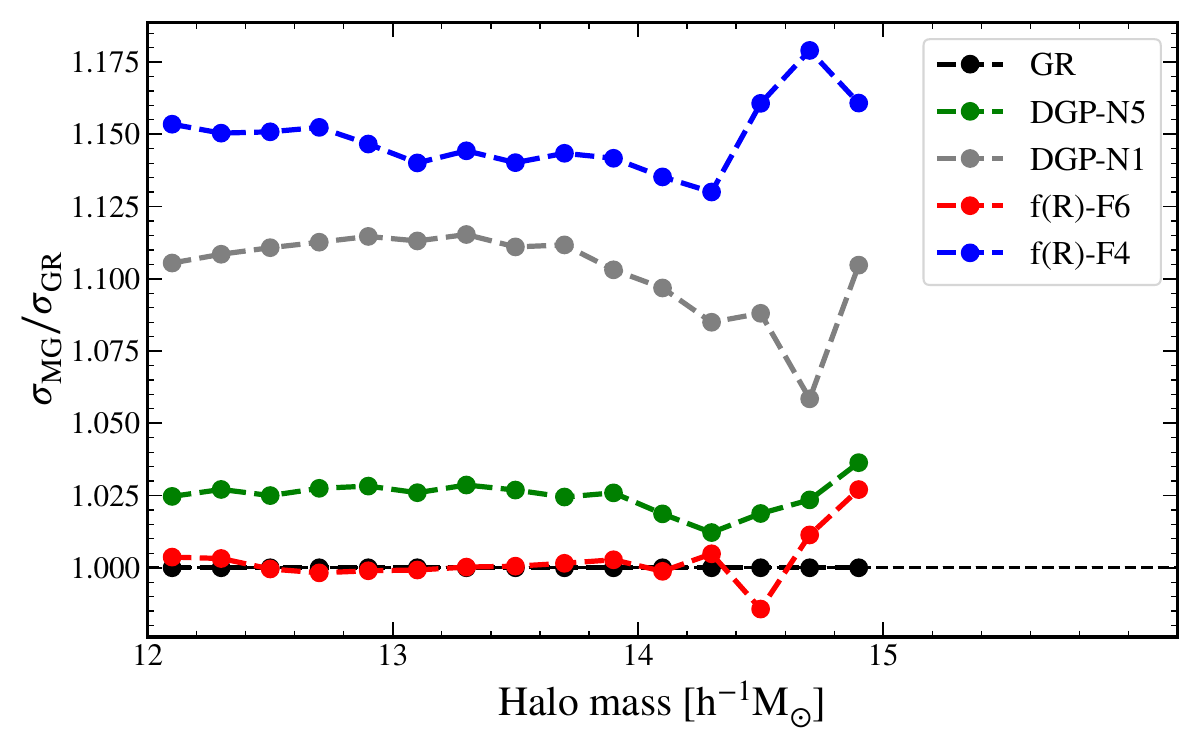}
    \caption{Velocity dispersion measured from the MG simulation compared with the GR result, as a function of halo mass at redshift $z = 0.55$. Different MG methods are respectively represented by distinct colors. In the calculation of the true value of $\gamma_{f}$ for four MG methods in Figure \ref{fig:MG}, the average value of the ratio $\sigma_{\rm{MG}}/\sigma_{\rm{GR}}$ is taken into account as a multiplicative factor to give an effective $\gamma_{f}$ parameter for these MG models. }
    \label{fig:velo_disp}
\end{figure}

Although our recovery tests on the MG simulations show that the constraint on $\gamma_{f}$ is not unity, we need to verify if it is close to the underlying true velocity field of the MG models.
This means that we need measure the velocity field of the host dark matter halos and compare the result with GR, i.e. with exactly the same meaning as $\gamma_{f}$ for GR-based model. In particular, we compute the velocity dispersion (VD) $\sigma_{v}$ of these halos as a function of halo mass and present the result in Figure \ref{fig:velo_disp}. The prominent feature is that stronger MG models have more substantial deviation from GR, which is consistent with our expectation. For the strongest modified gravity model, it exhibits a value that is $15\%$ higher than that in the GR model. In addition, the result shows some mass dependency especially at the very massive end. Since our analysis is for CMASS-like galaxies that are preferentially hosted by dark-matter halos of $13\Msunh$ and above \citep{White2011,Zhai2023b}, we simply compute the mean value of $\sigma_{\mathrm{MG}}/\sigma_{\mathrm{GR}}$ in the range of $12\sim15\Msunh$ and assume it as an effective correction for the velocity field of MG models. We can compute a more accurate correction by considering the probability distribution of host halo mass weighted by the HOD for each mock, but the difference is minor compared with the simple mean value correction.

With the VD correction for the extraction of the true value of $\gamma_{f}$ of each MG model, we calculated the residual of the recovery analysis compared to the true value of each modified gravity mock galaxy sample, as shown in Figure \ref{fig:MG}. 
As a byproduct, we also investigate the possible correlations between $\gamma_{f}$ and assembly bias parameters in the constraint on MG simulations, shown as different colors in the figure. 
The figure shows five panels for five HOD models we have considered, and each panel has a lower plot showing Bayesian evidence as a statistic for goodness of fit. As noted above, five cosmological parameters, namely $\Omega_{b}$, $h$, $n_{s}$, $\omega$, and $N_{\mathrm{eff}}$, are fixed in the analysis and the remaining cosmological parameters, $\Omega_m$ and $\sigma_8$, along with all the basic and extended HOD parameters detailed in Section \ref{subsec:HOD}, are allowed to vary. 

From the results, we can see that, first, our $\gamma_{f}$ model is able to measure an effective parameter of the velocity field of different MG models using the RSD effect through clustering measurement on non-linear scale. Compared with GR, the final result has some bias depending on the HOD models for the galaxy-halo connection, but none of the selected models has the remaining bias larger than 1$\sigma$. 
Second, the VD correction shows that the peculiar velocity field of the MG models is higher than GR and therefore the RSD effect is stronger, especially for models with more substantial deviation from GR. In this case, the recovered $\gamma_{f}$ parameter of the MG models is higher than GR.
Third, we also investigate the correlation between $\gamma_{f}$ and the galaxy assembly bias parameters. The result seems to show that adding these degrees of freedom to the model does not change the constraint on $\gamma_{f}$.

Beyond the study presented in Figure \ref{fig:MG}, we perform the tests by fixing $\gamma_{f}=1$ in our model constraint, we simply find that the $\chi^{2}$ and Bayesian Evidence is not able to yield a reasonable fit. This is not surprising since fixing $\gamma_{f}=1$ forces the model to be GR and thus it is not able to explain the MG theories.

In addition, recent studies seem to show that observational data prefer a lower growth rate parameter or amplitude of matter perturbation through analysis of galaxy clustering and galaxy lensing \citep{Hamana2020,Abbott2022,Zhai2023b,Hahn2023}. From this point of view, the MG models does not seem to be favored by observations since they preferably to have a higher amplitude of velocity field given the same cosmological evolution in the background. Note that our analysis is purely theoretical without careful examination of observational data but can indirectly support works that analyze survey data such as \citet{He2018}. 
On the other hand, we should also note that this is only valid for our particular analysis using clustering measurement at nonlinear scales, the correlation between MG models and assembly bias in terms of galaxy-halo connection models and analysis scales of interest is worth studying in future works.

\section{Conclusions and discussions}
\label{sec:Conclusion}

This research delved into two crucial aspects within the realm of cosmology at a redshift $z = 0.55$ and within the scales of $0.1\sim60\Mpch$ for galaxy clustering analysis: assembly bias and modified gravity. This scale of analysis couples both fundamental cosmology and galaxy formation physics, and therefore, unbiased and accurate cosmological measurement is challenging. Thanks to the development of emulator methodology, we are able to build an accurate model and investigate multiple impacts or correlations from a fully statistical point of view. Using $f\sigma_{8}$ and an empirical $\gamma_{f}$ parameter, the growth rate parameter as the key observable, we examine the impact of assembly bias in the clustering analysis and the possibility of measuring MG signals. Our findings can be summarized as follows:

\begin{itemize}
    \item Using the external environment as the secondary parameter, the assembly bias has a direct correlation with the cosmological measurements. A model without complete galaxy-halo connection prescription can yield biased cosmological inference and the offset is proportional to the strength of the clustering amplitude determined by the secondary parameters. 

    \item The single $\gamma_{f}$ parameter as a scaling factor of the halo velocity field is able to retrieve the Modified Gravity signature effectively.  The final measurement on the cosmological parameters may have a certain amount of remaining bias dependent on the HOD models, but the significance is not substantial given a CMASS-like sample.
  
    \item The behavior of galaxy assembly bias does not have a strong correlation with the MG effect in terms of the RSD measurement, but it is worth more thorough investigation in future works to isolate one from the other.
\end{itemize}

In this paper, we used only one redshift snapshot from multiple simulation suites, and we expect the above conclusion to hold in a wider redshift range with different number densities of galaxy samples. The analysis also demonstrates the power and cosmological information hidden at non-linear scales. With more summary statistics such as higher-order correlation function or galaxy-galaxy lensing, we can anticipate another leap in the significance of the measured signal. The current emulator-based methodology is ready to be applied to ongoing surveys such as DESI, which has collected a large amount of data from LRG, ELG, and QSO \citep{DESICollaboration2022,Raichoor2023,Zhou2023,Yeche2020}. With increased survey volume and number density, it can provide an ideal framework to exploit and retrieve information as we present here \citep{Dawson2022}, including but not limited to galaxy assembly bias and modified gravity. 

This study's emulator model is based on the HOD model, focusing on second-order properties with a given galaxy number density. The Conditional Luminosity Function (CLF), which directly impacts galaxy number density in different halo mass, can be added to enrich the model. Future research may focus on integrating the latest CLF measurements from DESI \citep{Yang2003,Yang2009,Wang2024} into the emulator. 
Meanwhile, the subhalo mock catalog identified by \textsc{HBT+} code~\citep{han2012resolving,han2018hbt+} from Jiutian simulation suite, a high-resolution N-body simulation which is designed to satisfy the scientific requirements for the Chinese Space-station
Survey Telescope (CSST, \citealt{Zhan2011,Gu2024}), can be also applied in the galaxy-halo connection confidently.
This integration has significant potential to improve the accuracy and reliability of the model. Moreover, it is likely to offer more profound insights into the fundamental physical processes governing the universe, thereby contributing to a more comprehensive understanding of cosmic phenomena.
Upcoming research should focus on enhancing emulator models to better address the complex dynamics of cosmology and galaxy formation, as well as to enhance the accuracy of cosmological parameter estimations. This will, in turn, advance our knowledge of the structure and evolution of the universe.

\section*{Acknowledgements}

This work is supported by the National Key R\&D Program of China (2023YFA1607800, 2023YFA1607804, 2023YFA1605600), the National Science Foundation of China (No. 12373003), “the Fundamental Research Funds for the Central Universities”, 111 project No. B20019, and Shanghai Natural Science Foundation, grant No.19ZR1466800. We acknowledge the science research grants from the China Manned Space Project with Nos. CMS-CSST-2021-A02 \& CMS-CSST-2025-A04, and Yangyang Development Fund.

The computations in this paper were run on the Gravity Supercomputer at Shanghai Jiao Tong University.
We additionally made use of softwares: Numpy \citep{van2011}, Matplotlib \citep{Hunter2007}, SciPy \citep{scipy,Virtanen2020}, George \citep{Ambikasaran2015}, Corrfunc \citep{Sinha2020}, and MultiNest \citep{Feroz2009}.

\begin{appendix}
\setcounter{figure}{0}
\setcounter{table}{0}
\renewcommand{\thefigure}{\Alph{figure}}
\renewcommand{\thetable}{\Alph{table}}

\section{Impact of assembly bias on the galaxy two point correlation function}
\label{sec:Deviation_2PCF}

\begin{figure*}[htb]
    \centering
    \includegraphics[width=0.8\textwidth]{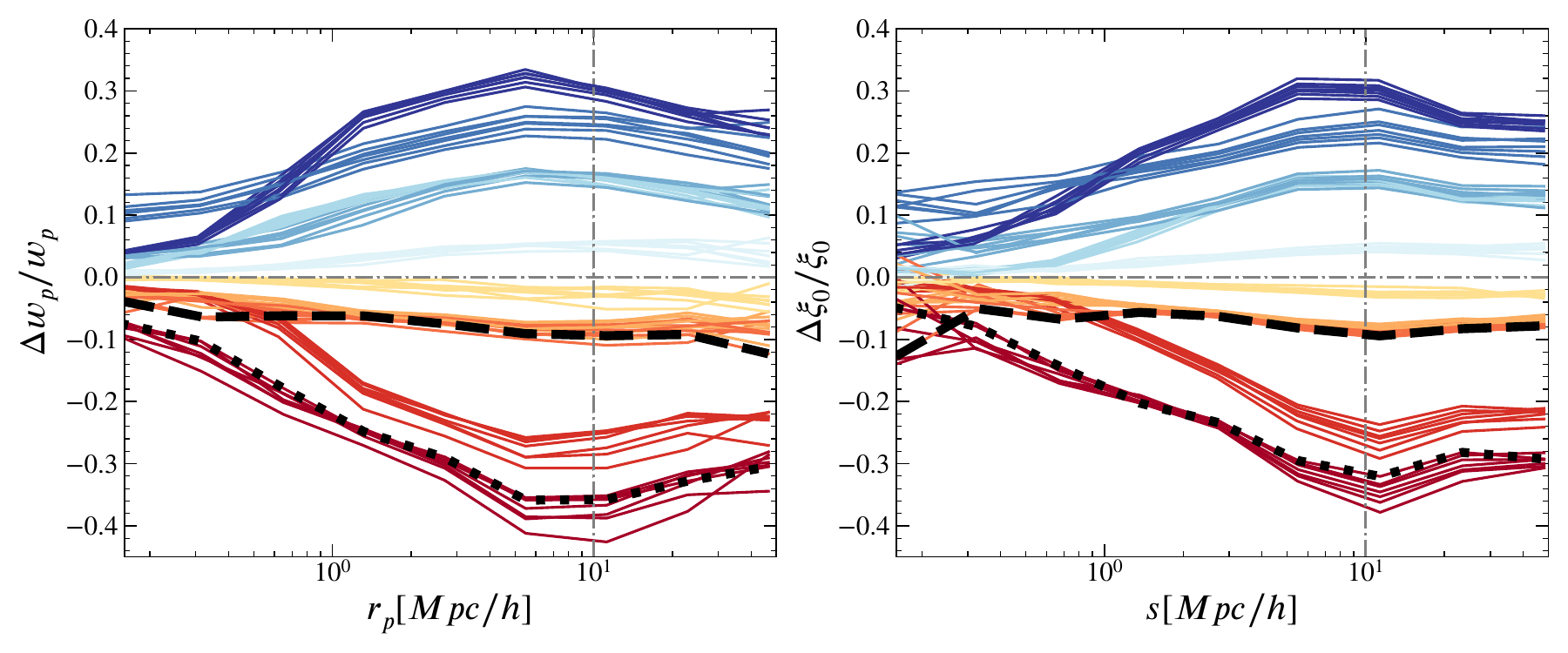}
    \caption{fractional change of clustering measurement (projected $w_p$ for left panel and monopole $\xi_0$ for right panel) for galaxy mocks with and without assembly bias effect. Different color are used for 10 different HOD models, and 7 test cosmology models for each HOD model have the same color. The maximum change occurs at $\sim10\Mpch$ for most of the models we have considered. The black dashed and dotted lines are associated with the left and right panels in Figure \ref{fig:AB_contour}, respectively. }
    \label{fig:AB_10MPC}
\end{figure*}

The impact of assembly bias on galaxy clustering can depend on both the cosmology and the HOD model. In Figure \ref{fig:AB_10MPC}, we show the fractional change of the galaxy clustering for mocks with and without environment-dependent assembly bias, for ten HOD models and all seven test mocks from the Aemulus suite. Different HOD models are represented by different color intensities, i.e. for each HOD model, there are seven cosmology models denoted by the same color.
We can see that within each HOD, the change in cosmology is not significant.
On the other hand, the fractional change has a diverse distribution for different HOD models, indicating that the modeling of galaxy-halo connection is more dominant than cosmology at non-linear scales. Moreover, we can see that the offset is maximized around $10\Mpch$. Therefore, we can use the measurement on this scale to approximately quantify the strength of assembly bias as elaborated in Section \ref{subsec:AB}.

\section{Galaxy clustering for mocks}
\label{sec:2PCF}

\begin{figure*}[htb]
    \centering
    \includegraphics[width=1\textwidth]{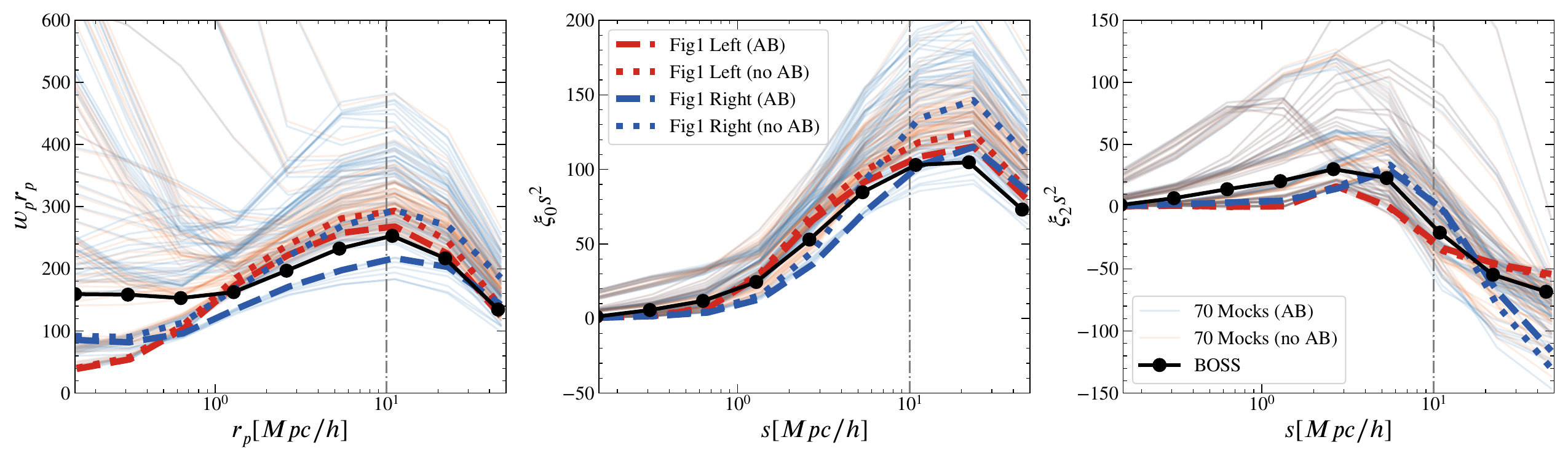}
    \caption{Galaxy clustering results for mocks with and without assembly bias. Two illustrative cases from Figure \ref{fig:AB_contour} are represented by thick dashed and dotted lines in two distinct colors. The ensemble of 70 mock (7 cosmologies × 10 HOD models) catalogs is depicted with thin light lines, while black lines denote BOSS dataset measurements for comparison.}
    \label{fig:AB_clustering}
\end{figure*}

\begin{figure*}[htb]
    \centering
    \includegraphics[width=1\textwidth]{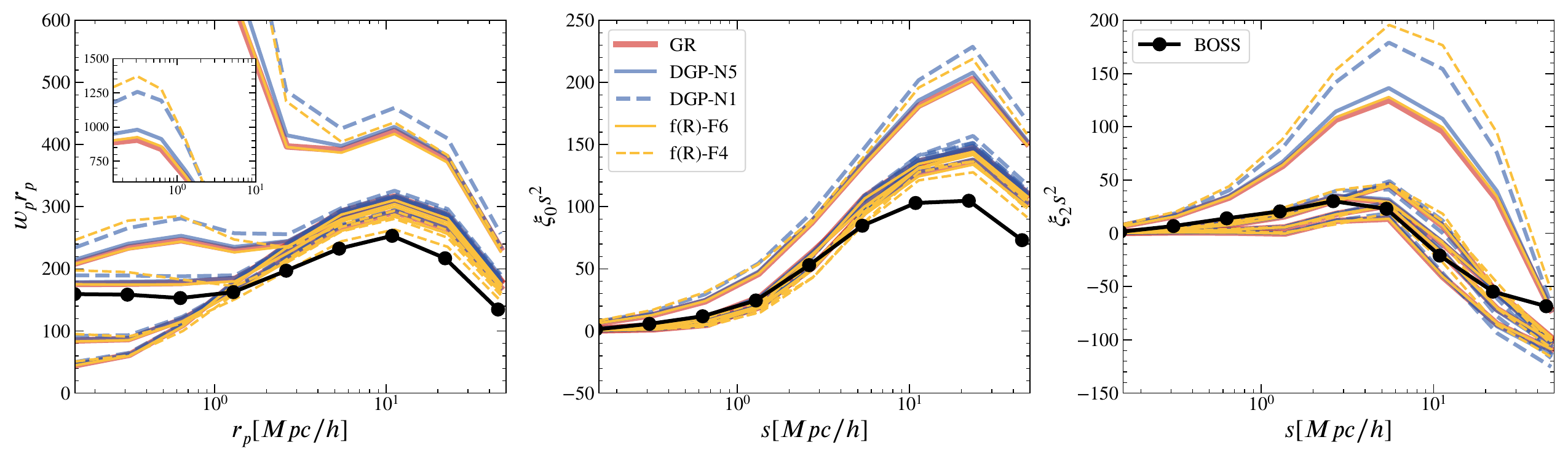}
    \caption{Galaxy clustering results are presented for modified gravity mock catalogs across 5 HOD models sharing identical cosmological parameters which correspond to the five panels in Figure \ref{fig:MG}. GR and four modified gravity methods are distinguished via distinct line styles. For comparison, measurements from the BOSS dataset are overlaid as black lines.}
    \label{fig:MG_clustering}
\end{figure*}

In this section, we present the clustering results for the mocks used in the analysis. In Figure \ref{fig:AB_clustering}, we show the galaxy clustering results ($w_p$, $\xi_0$, and $\xi_2$) for two types of mocks: those with and without assembly bias for Section \ref{subsec:AB}. Notably, the two illustrative cases displayed in Figure \ref{fig:AB_contour} are highlighted by thick dashed and dotted lines in two distinct colors. The red one corresponds to the left panel of Figure \ref{fig:AB_contour}, and blue one corresponds to the right panel. Additionally, the galaxy clustering results for all mocks involved in Figure \ref{fig:AB_deviation} are depicted with thin lines in the plot. Figure \ref{fig:MG_clustering} presents results for the five different HOD models shown in Figure \ref{fig:MG} for the MG analysis in Section \ref{subsec:MG}, where distinct line styles represent different modified gravity methods. To enable comparison with real galaxy clustering measurements, we overlay the BOSS dataset measurements as black lines in the figures.

\end{appendix}
\clearpage
\bibliographystyle{aasjournal}
\bibliography{main}

\end{CJK*}
\end{document}